# Experimental analysis of a multistage water desalination system utilizing an evacuated parabolic solar trough with a solar tracker

Abbasali Davani; Mehdi Kashfi; Vahid Mozafary

Bahonar University of Kerman

**Abstract**:

The paper describes the experimental investigations of the performance of a multi-stage water desalination still connected to a solar parabolic trough (solar energy concentrator) with focal pipe and simple heat exchanger (serpentine). The receiver length is 2 m and has a diameter of 6 mm. A diluted solution of water-antifreeze is selected as working fluid. This solution is flowing as cycle through the focal pipe and serpentine. The solution in the heat exchanger is heated during the sunlight by means of solar energy concentrator (solar parabolic trough). The parabolic trough is tracking the sun using simple mechanism to collect the highest amount of solar radiation by optimum angle. The receiver is covered with transparent borosilicate glass to avoid the heat transfer by convection to the surrounding (i.e. the greenhouse effect is utilized) and the space between the glass and tube is evacuated. Aluminum serpentine of the heat exchanger is installed in the bottom of the still. It is attached extremely with thermally insulated pipelines through the trough. The fluid is forced flow using small pump which is powered by PV system. The multi-stage solar still water desalination system was designed to recover latent heat from evaporation and condensation processes in 2 stages. The results of tests demonstrate that the system produces about 3.6 kg of fresh water for only 5 hours from 10:00 to 15:30.the The system components were fabricated and the overall system was assembled in International Center for Science & High Technology & Environmental Science, Mahan, Iran.

**Keywords**: Solar Energy; Desalination; Solar Parabolic Trough; Renewable Energy, Multi-staged still

1. Introduction

The world's population is augmenting, and fresh water has always been one of the most vital requirements for life to continue on the earth. However, while water covers almost 75% of the earth's surface, only 3% is fresh water from several sources, and from this finite quantity a small portion of it can be used for drinking. Hence**,** water treatment is usually necessary. It is proved that the most efficient way for providing fresh water is desalination of brine and/or seawater. But, this method is quite energy intensive, and needs lots of oil and wood and the operational cost would be high; so a cost-effective alternative would be using solar energy which is safe, free and renewable for the desalination process and in the areas with greater fresh water shortages like the southern Mediterranean and the Middle East, the amount of this nature benign kind of energy is satisfyingly during four seasons of the year [1]

As mentioned before, one of applicable ways to produce fresh water in remote areas is to use thermal solar energy to desalinate saline water [2] and an example of this practical and yet simple technology is conventional basin solar stills which have a relatively large footprint area But, this kind of solar stills have several disadvantages like low thermal efficiency, low efficiency because of extreme heat losses, and developing reduction in efficiency over the utilization time due to the scaling and sedimentation of salt impurities [3]. It is seen that the maximum thermal efficiency for basin solar stills is regularly about

25%, and the average distillate output capacity is 1.5–3.0 kg/$m^2$/day [4]. Investigation on application of other types of solar stills like compact multi-stage tray solar stills which use flat-plate solar air and water collectors, has been carried out by groups of researchers too. For instance, the efficiency of a closed loop solar desalination unit with humidification and dehumidification was examined by Yuan and Zhang in Ref. [5]. The theoretical and experimental result of examinations on a small-scale barometric desalination system which was powered by solar energy was described by Eames et al. [6]. Jubran et al. [7] undertook a numerical modeling of a multi-stage solar still with an expansion nozzle and heat recovery operating in a steady state regime. Adhikari et al. [8] described theoretical studies on a multi-stage tray solar still for steady state conditions. At first a mathematical model of a three-stage unit and their theoretical results were presented by the authors and then they were validated by means of an experimental electric heater energy source. Abakr and Ismail [3] and Toyama et al. [9] also examined multi-effect solar stills. Schwarzer et al. in [10] used a 6-stage square still with a 0.64 $m^2$ area as a solar desalination unit connected to 2 $m^2$ flat-plate collectors. After performing numerical simulations it was seen that on the basis of the insolation intensity of 4.8 kW/$m^2$/day the distillate rate would be 25 kg/$m^2$/day. Although this specific productivity is considered extreme but practically, reaching such productivity is never occurred. A maximum production capacity of 4.6 kg/day was reached after Badran et al. [11] examined a basin still with a 1$m^2$ area joined with a 1.34 $m^2$ flat plate collector and the overall efficiency was 27%. Abu-Jabal et al. [12] built and tested a three-stage desalination system in the Gaza Strip. The distillation unit which was coupled with a solar collector and photovoltaic cells has dimensions of 9 m length, 3.2 m width and 2.3 m height and the air was evacuated in the still. The highest and the lowest water production capacity were 204.5 kg/day and 5.03 kg/day in July and January respectively. The results of a system modeled using heat and mass transfer equations were described by Garg et al. [13]. In their model, the desalination unit was supplied by hot water continuously in 24 h which was provided from a solar system. According to the obtained data, the relation between the production rate of distilled water and the temperature of water in the humidifier was linearly proportional.

Analytical and experimental analysis of a multistage solar still with stacked array of distillation trays was investigated by Fernandez and Chargoy [14]. Bottoms were w-shaped and they act as condensers for the below trays. Two major conclusions can be resulted from their work; first that desalinating of seawater in a multi-staged still is convincing and second that the overall efficiency of the still would be affected as the flow of steam bypasses the condenser. A computer simulation model for performance of a multistage-stacked tray solar still in steady state regime using the modified mass and heat transfer relationships was presented by Adhikari et al. [8] and the results obtained from the model was in a good agreement with the experimental results.

Solar-driven distillation is attractive due to two factors. First, it is economical aspect of this type of distillation since desalination and distillation specifically are processes with incentive need of energy, with the major expenses of required operating energy and if one could drive a process of desalination using the abundant energy provided by the sun, it would be ideal. The other reason which has made this type of desalination distillation worthwhile is geographical considerations. Most desert and dry regions that are in extreme need of fresh water are those which solar irradiation is high, and where there is absence of fresh water solar irradiation is abundant, it would be very tempting to use solar energy as providing energy for desalination process. But it is significant to bear in mind that although solar irradiation releases high amounts of energy, effective collection is usually occurring at low temperatures (below 100°C) due to low concentration of solar irradiation. [15] An ideal solution to harness this energy for desalination would be an apparatus that would concentrate the sun's energy all day long and make use of this plentiful energy efficiently. M.I. Ahmed et al. [16] used a multistage evacuated solar still and the total daily yield was found to be almost three times of the maximum productivity of the basin-type solar still but evacuating the still is not an easy and cost-effective way. Furthermore, the maintenance will not be accomplished easily, so an alternative way to have both high production and low cost would be using of a solar collector and a multi stage solar still that uses the solar energy efficiently. In this paper use of a solar parabolic trough with a solar tracker and horizontal focal parabolic concentrator is used to provide energy for a 2-stage solar still.

## 2. Description of System

The experimental set-up solar desalination systems is composed of a 2 stage solar still and the solar parabolic trough and a PV panel in order to provide power for motor and pump Figs.1 show the schematic diagram and photography pictures of the test rig with which was built at International Center for Science & High Technology & Environmental Science [17].

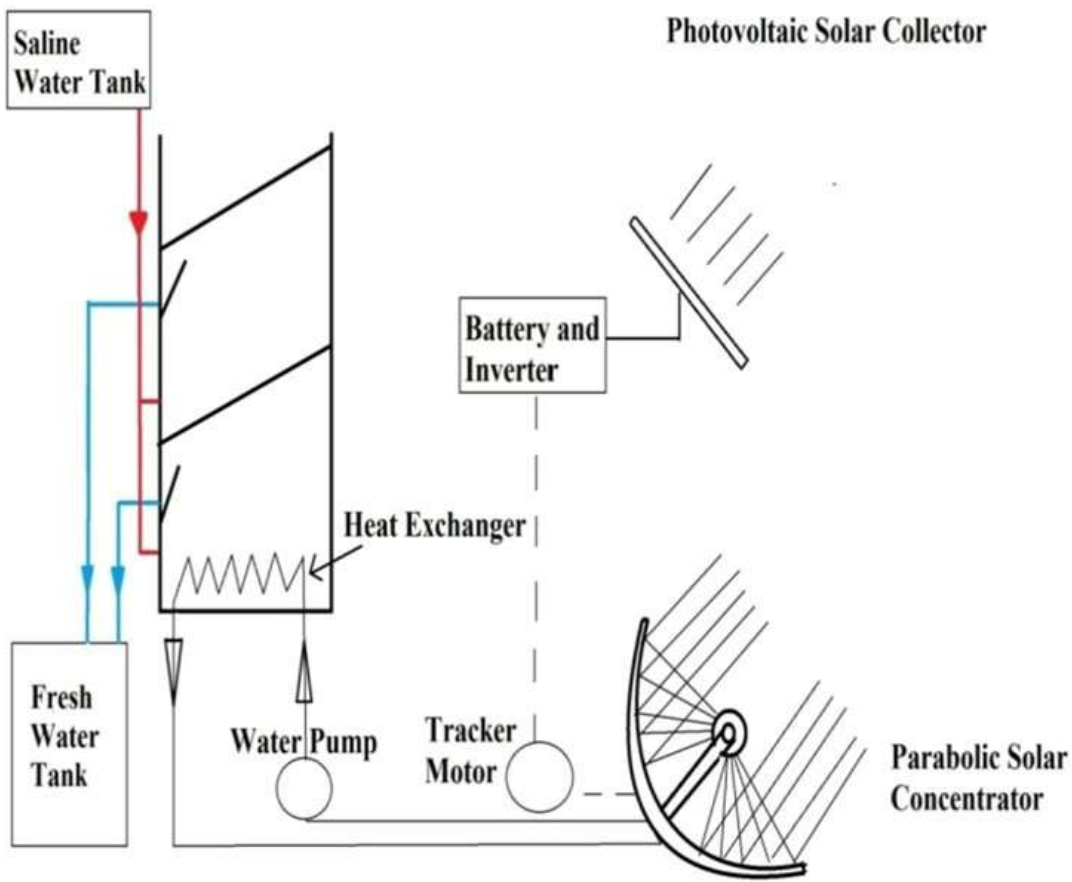

Fig.1.1 Schematic diagram of presented Solar still

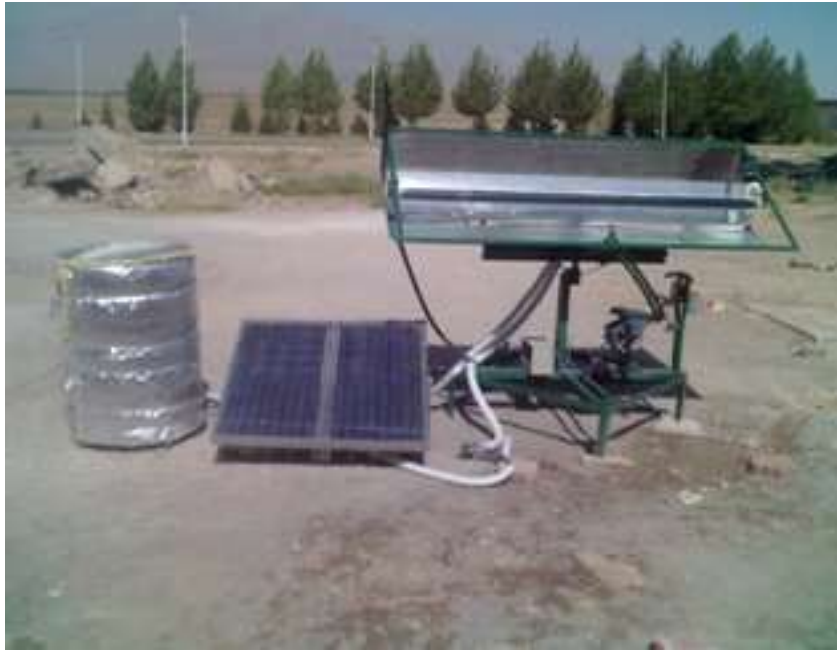

Fig.1.2 the experiment test rig

The experiment is carried out under the case study of forced water flow at July 2010. The data of the typical climatic conditions of the experimental Kerman site at the selected days are obtained from data processing center of meteorological organization (IRIMO) [18], station Kerman and (Figs. 5–8). In the following, different elements of the system are described and characterized

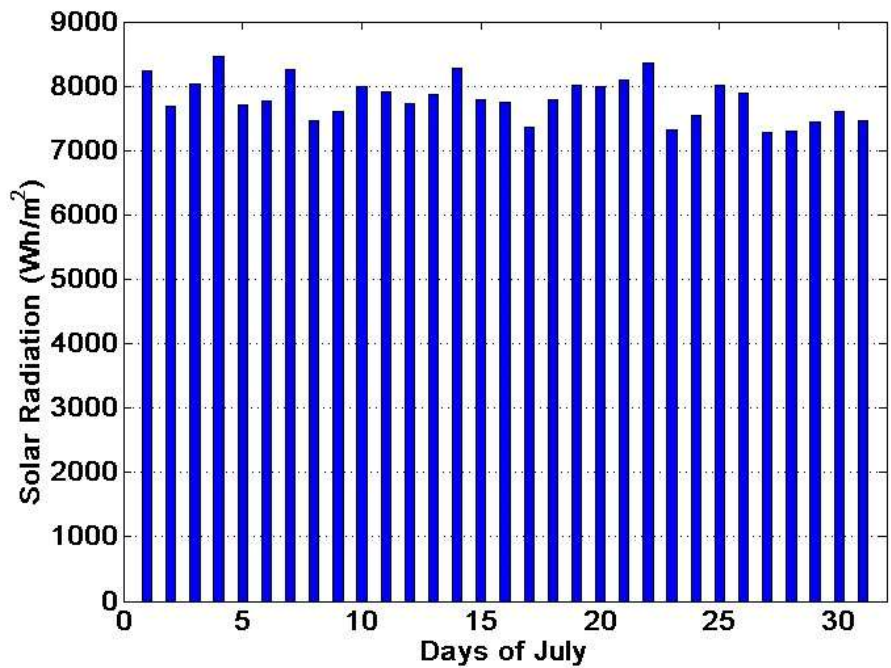

Fig.2. Average daily solar radiation collected data from Kerman in July 2009

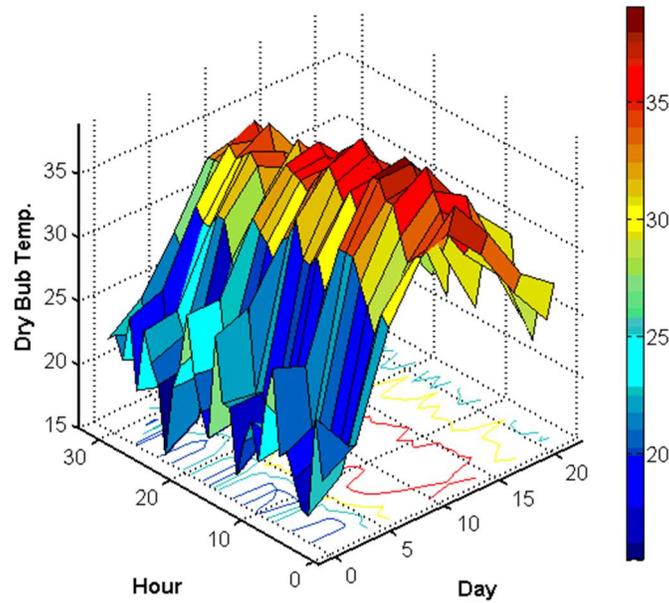

Fig.3. Daily collected ambient temperature data from Kerman in July 2009

2.1. *Parabolic concentrator and its optical analysis*

The parabolic trough is solar concentrator, reflector and collector. Fig.4 shows the used parabolic solar concentrator which has the length of 2 m and the span of 0.8 m with the focal length of 0.45 m and the receiver is a 60 mm aluminum pipe that is painted with black color to absorb the highest amount of incident solar radiation during sunlight. The receiver then is put in a tubular borosilicate glass and the endings are perfectly fitted so that the air between the receiver and glass can be evacuated in order to prevent convection and conduction heat transfer from receiver to ambient and the whole apparatus is designed and manufactured in International Center for Science & High Technology & Environmental Science, Mahan, Iran.

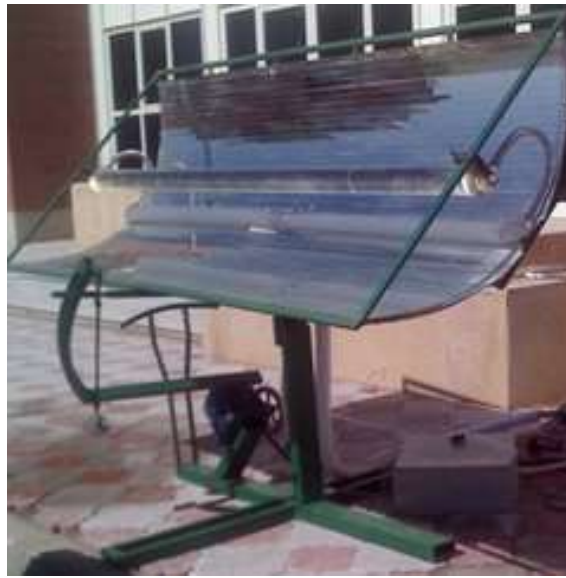

Fig.4.The solar parabolic trough used for providing energy for still

The optical analysis of the present parabolic concentrator is presented.

To describe the amount of light energy concentration achieved by a given collector the term "*concentration ratio*" is used. Two different definitions of concentration ratio are in general use, Optical and geometrical concentration ratios. They are defined briefly here so that the terms may be used. These ratios are defined as the following [19]

*Optical Concentration Ratio,* $(CR_o)$ is the averaged irradiance (radiant flux) $(I_r)$ integrated over the receiver area $(A_r)$, divided by the insolation incident on the collector aperture.

$$CR_o = \frac{\frac{1}{A_r} \int I_r dA_r}{I_o} \quad (1)$$

*Geometric Concentration Ratio,* $(CR_g)$ is the area of the collector aperture $A_a$ divided by the surface area of the receiver $A_r$

$$CR_g = \frac{A_o}{A_r} \quad (2)$$

$CR_o$ relates directly to lens or reflector quality; but lots of collectors have receivers with the surface area larger than the concentrated solar image. Parabola is defined as the locus of a point that moves so that its distances from a fixed point and a fixed line would be the same. The fixed line is called the *directrix* and it is shown on Figure 1 and the fixed point *F*, is the focus. Note that the *RD* and *FR* have the same length. *Axis of the parabola* is the line passing through the focus *F* and is perpendicular to the directrix. *Vertex* is the point in which parabola intersects its axis, and it is exactly midway between the directrix and focus.

Equations of parabola are given as [8]:

$$y^2 = 4fx \quad (3)$$

where *f*, the focal length, is the distance *VF* from the vertex to the focus. Shifting the origin to the focus as is often done in optical studies, with the vertex to the left of the origin, the equation of a parabola can be written as

$$y^2 = 4f(x + f) \quad (4)$$

And in polar coordinates we have:

$$\frac{\sin^2 \theta}{\cos \theta} = \frac{4f}{r} \quad (5)$$

where *r is* the distance from the origin and $\theta$ is angle from the *x*-axis to *r*. Note that parabola's vertex is at the origin and symmetrical about the *x*-axis

The distance from the focus *F* to the curve is called the parabolic radius $r_p$ (m) is and can be computed as following:

$$r_p = \frac{2f}{1 + \cos} \quad (6)$$

where $\psi = 2P$ is careful inspection of the geometry described in Fig 5. it is more useful to define the parabolic curve with the origin at $F$ and in terms of the angle $\psi$ in polar coordinates with the origin at $F$

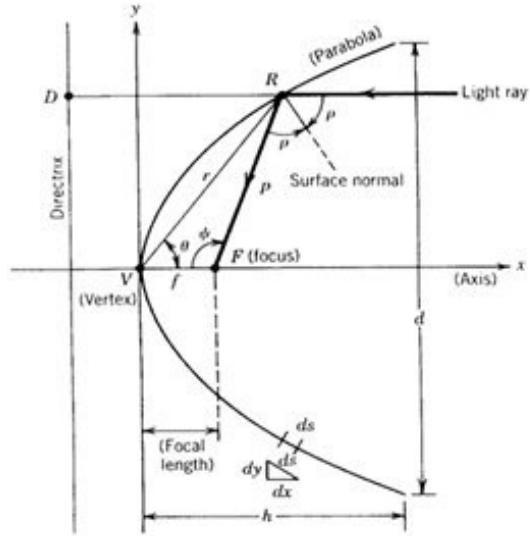

Fig.5.Profile of present concentrator

While The general expressions given so far for the parabola define a curve infinite in extent, solar concentrators use a truncated portion of this curve. The extent of this truncation is usually defined in terms of the ratio of the focal length to aperture diameter $f/d$ or the rim angle $\Psi_{rim}$ or the focal length $f$ or aperture diameter $d$ are linear dimensions used to specify the scale (size) of a curve. It can be seen in Fig.6 which shows various finite parabolas having a common focus and the same aperture diameter.

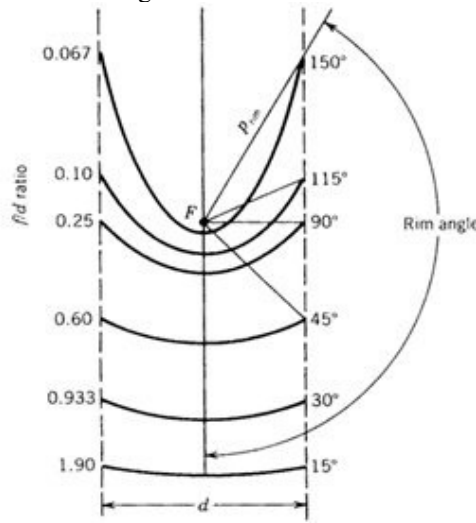

Fig.6. Segments of a parabola having a common focus $F$ and the same aperture diameter.

It is apparent that a as the rim angle reduces parabola tends to be flat and its focal length would be long compared to the aperture diameter. The height of the curve, $h$ may be defined as the maximum distance from the vertex to a line drawn across the aperture of the parabola once a specific portion of the parabolic curve has been selected. The height of the parabola can be written in terms of focal length and aperture diameter as:

$$h = \frac{a^2}{16f} \tag{7}$$

In a same way, the parabola dimensions may be used to find the rim angle $\Psi_{rim}$:

$$\tan \Psi_{rim} = \frac{1}{(\frac{d}{8h})-(\frac{2h}{d})} \tag{8.1}$$

$$\tan (\frac{\Psi_{rim}}{2}) = \frac{(\frac{f}{d})}{2(f/d)-(\frac{2h}{d})} \tag{8.2}$$

The arc length $s$ is another property of the parabola which may be used in understanding solar concentrator design and it can be found for a particular parabola from Equation (3) by integrating a differential segment of this curve and applying the limits $y = d/2$ and $x = h$ as showed in Fig.5, the result is

$$s = [\frac{d}{2}\sqrt{\frac{4h^2}{d}+1}] + 2f \ln[\frac{4h}{d}+\sqrt{\frac{4h^2}{d}+1}] \tag{9}$$

where $d$ and $h$ are the distance across and the distance from the vertex to the aperture of the parabola, respectively as shown in Fig.1. The cross sectional area of the space enclosed between a parabola and a line across its aperture and normal to the axis is given by

$$A_x = \frac{2}{3}d.h \tag{10}$$

Since a parabolic trough has a line focus, in order to maintain focus it must be tracked about its linear axis. The proper tracking angle is defined by the orientation of the trough relative to the sun's position. Ref. [18] presents analytical expressions need to find proper tracking angles for solar parabolic troughs.

### 2.2. multi stage still

The water desalination system is a two-stage tray laboratory prototype cylindrical still made of 2mm galvanized iron sheet coupled with a solar parabolic trough. Each stage of the still has a diameter of 600 mm and the heights of stages are 450 mm and 350 mm respectively containing angled trays. The main portion of the saline water is contained in the first (bottom) stage which has the largest volume and has the water depth of about 40 mm. a serpentine aluminum tubular heat exchanger is installed horizontally and it is covered by water at the bottom of the still and has outer diameter of 1 cm and total length of 14 m. a closed circuit of water-anti freezer which is circulated in the heat exchanger and the evacuated solar collector by means of an electrical pump is formed by connecting inlet and outlet of collector to the outlet and inlet, respectively, of heat exchanger. By varying the pump's speed, we can control the flow rate in the circuit and by using a high temperature flow meter it can be measured. The second stage of the still is same in design and an angled tray has covered it's top. In both stages the trays are identical in design and they shaped in oval aluminum sheets with slope of 10 °. Trays are chosen to be aluminum with thickness of 1 mm in order to have a high thermal conductivity and also their surfaces were grooved in order to maintain drop wise condensation because the surface was coated with a substance that inhibits wetting and film condensation could not be maintained  A small  bowl-shaped trough is installed in both stages just below the ridge of the slopping tray and it's lower end is connected to the distillate pipe outlet so that the distillate water collects initially in the trough and then to the metering cylinders. A special silicone rubber sealant strips are placed between the installation surfaces in order to prevent the escape of the water vapour from the system. Brackish water is fed into the still from inlet points at each stage and it can be controlled by adjusting the height of a hot water float valve. When the level of saline water reaches the desirable height in both stages then the valves will close automatically. The process that occurs in the multi-effect still takes place in different phases. First, the fluid which is circulated in the ''heat exchanger- solar collector'' loop uses the energy absorbed by the solar collector to increase its temperature. Then, the circulation of hot fluid in the heat exchanger causes the heat to transfer to brine in the first stage of the still and as the temperature of saline water in this stage increases vapur starts to produce. Vapour then condensates on the slopping surface of the above tray due to the temperature difference of the saline water between the stages and droplets of the condensate which are formed on the surface of the above tray flow towards the edge of the slope due to gravity and collected in the trough. And finally, the distillate water flows from the trough to metering cylinders. The temperature of saline water in the following stage gradually increases utilizing the latent heat of condensation and hence, the process of evaporation occurs in the second stage. A 50 mm insulation of glass wool with thermal conductivity of 0.04 (W/mK) is also used to reduce heat losses of the solar still to the ambient. The temperature of the saline water in both stages and the temperature of the circulating fluid in the heated circuit are measured by means of different thermocouples installed in different points of apparatus. The connecting pipes between serpentine heat exchanger and solar collector were well thermo-insulated to reduce thermal losses.

## 3. Thermal analysis of the present solar desalination system

As mentioned before, the brackish water in the first stage is heated by circulation of diluted solution of water-anti freezer from the solar collector to the heat exchanger. Water in the second stage is heated using both sensible heat of vapour coming from first stage and the latent heat of condensation.

The following mathematical model for the still is derived base on below assumptions:
1. Non-condensable gases released from water have no effect in throughout process.
2. The distillate output and the amount of evaporated water in both stages are the same.
3. The temperature of the condensation surface and the distillate leaves the desalination system are equal.

The analytical model proposed is similar to that derived in Ref. [8-19]. It is lumped-parameter set of ordinary differential equations as mass and energy conservation equations written for both stages of the still. Fig.7 presents a schematic energy balance diagram of the still.

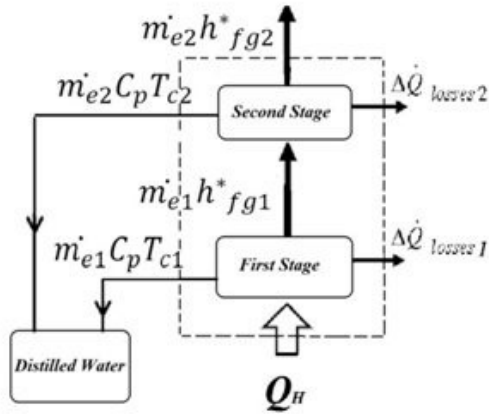

Fig.7. Flowchart showing energy flow through the system.

First, we express equations of energy conservation as follows:

In the first stage we have:

$$\dot{Q}_H - \dot{m}_{v1}\left(h^*_{fg1} + c_p T_{u1}\right) = M_{s1} C_p \frac{dT_{s1}}{dt} + \Delta \dot{Q}_{L1} \qquad (11)$$

And in the second stage of still,

$$\dot{m}_{v1}h^*_{fg1} - \dot{m}_{v2}(h^*_{fg2} + c_p T_{u2}) = M_{s2}C_p\frac{dT_{s2}}{dt} + \Delta\dot{Q}_{L2} \qquad (12)$$

where $\dot{Q}_H$, $\dot{m}_v$, $C_p$, $h^*_{fg}$, $\Delta\dot{Q}_L$, $M_s$, $T_u$, and $T_s$ are collector's output heat (W), vapour mass flow rate in stage (Kg/s), specific heat of water (J/KgK), refined value of latent heat of vaporization of water at the condensation surface (J/Kg), rate of heat losses in the stages of sill(W), mass of water bed in stages, upper surface temperature and water bed temperature in the stages(°c), respectively.

In equation (11)

$$\dot{Q}_H = \dot{m}_{sc}C_{p\,solution}(T_{SC\,inlet} - T_{SC\,outlet}) \qquad (13)$$

where $\dot{m}_{sc}$, $T_{SC\,inlet}$ and $T_{SC\,outlet}$ are mass flow rate of the fluid in the solar collector (kg/s), Solar collector inlet and outlet temperatures (°c) and here $C_{p\,solution}$ is specific heat of diluted solution of water and Propylene glycol. The solvent is added to increase boiling point of water.

For each stage, we can write equation of mass conservation as:

$$\frac{dM_{si}}{dt} = -\dot{m}_{vi} \qquad (14)$$

Note that *i* is number of each stage.

Knowing the differences in temperatures of the brine in stages and ambient surroundings and the heat conductivity coefficient of the glass wool, heat losses were calculated and they found to be low.

Refs. [20,21] proposed the relations between the water's refined latent heat of vaporization on the current temperature and the latent heat's magnitude and here, the same procedure is used:

$$h_{gfi}(T) = 1000 \cdot [3161.5 - 2.4074(T + 273)] \qquad (15)$$

$$h^*_{fg} = h_{gfi} + 0.68 \times C_{pi}(T_{si} - T_{ui}) \qquad (16)$$

Water's heat capacity can be defined as a function of its temperature as proposed in Ref. [22]:

$$C_{pi} = 1000 \times [4.2101 - 0.0022 \times T_{si} + 5 \times 10^{-5} \times T_{si} - 3 \times 10^{-7} \times T_{si}] \qquad (17)$$

According to second assumption the flow rate which distillate comes from stage can be same to the rate in which the brackish water evaporates in the corresponding stage and it can be computed as

$$\dot{m}_e = \frac{(T_{si} - T_{ui})h_{ewi}A_{si}}{h_{gfi}} \qquad (18)$$

where $h_{ew}$ is the coefficient of evaporative heat transfer and it is determined as proposed in Ref. [23]

$$h_{ewi} = 0.884 \times \left[(T_{si} - T_{ui}) + \frac{(P_{si} - P_{ui})(A_{si} + 273)}{286.9 \times 10^3 - P_{si}}\right]^{\frac{1}{3}} \qquad (19)$$

$P_{si}$ and $P_{ui}$ are partial vapour pressures and can be computed as proposed in Ref. [23]

$$P_i = e^{\left(25.317 - \frac{5144}{T_i + 273}\right)} \qquad (20)$$

And finally we can define the still's distillation efficiency as

$$\eta_{distll} = \frac{\sum_{i=1}^{2} \dot{m}_{ei} h_{fgi}}{\dot{Q}_H} \tag{21}$$

## 4. Experiments procedure

The water volumes in the first and second stages were 15.7 and 12.9 L, respectively. The water depth in the angled trays varied from 0 to 90mm. The system was tested when the top of the second stage was left open i.e. it was not isolated so that air could flow on upper tray. The system was tested in a typical summer day in middle regions of Iran. Thus the information on the variation of the solar radiation on 8 July 2010 is shown in Fig.8

Basically, a parabolic trough must track about its linear axis so that when the sun's rays are projected onto the plane of curvature, they are normal to the trough aperture therefore a 2 axis sun tracker which tracks the sun all day long by means of a timer is used. The tracker is developed in a way that during the day the trough follows a curvature on a hypothetical sphere that is similar to curvature of sun in sky. The timer is set to move the tracker for 1.5 second every 5 seconds The mass flow rate through the collector was kept at 4.5 L/min. water that condensed during the experiments was collected into two metering cylinders and the amount of the condensate was measured after the 5 h period (10 A.M to 3 P.M) Measurements of the ambient temperature, wind speed, relative humidity, and radiation were conducted using a data logging system, which continuously stores the readings of each one of the different variables at 15 minutes intervals. Figs. 8-9 show the average wind speed and relative humidity during the test period.

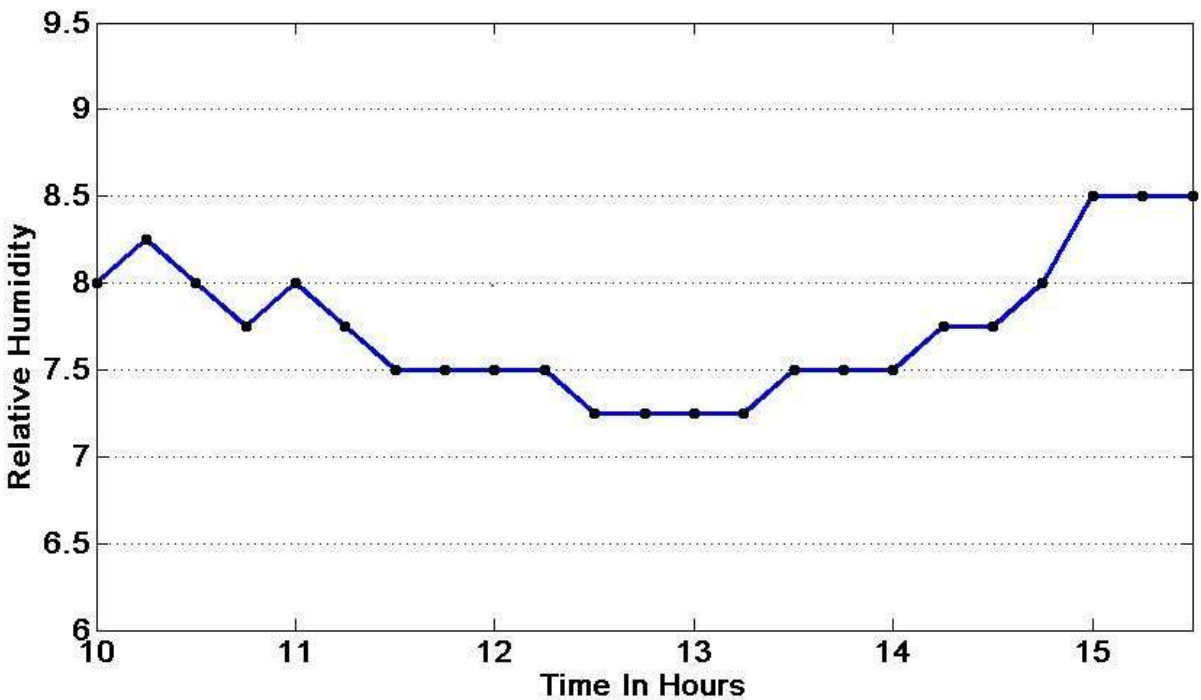

Fig. 8.Hourly relative humidity in the test day

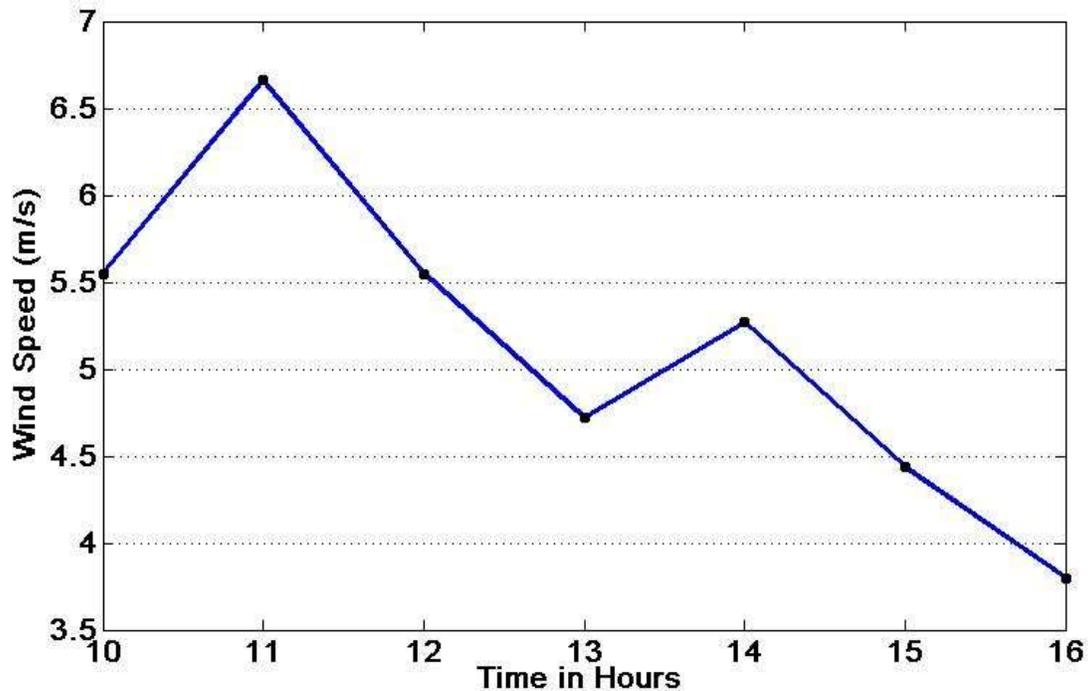
Fig.9. Average wind speed in the test day

## 5. Results and discussions

Figs. 10–14 present radiation on the 30 ° surface collected in test region, temperature variation of the components, and variation of accumulated fresh water for the solar desalination systems under case study. According to the hourly variation of the solar radiation, $I$, for the solar desalination systems at date day July 8/2010 the results have been illustrated. The ambient temperature,$T_{amb}$, varied and have peak values around the noon interval (12–2) .This variation and the hourly variation of temperatures of water in stages are plotted in Figs. 11 and 12. It is noticed, as time goes on, temperature increase, but those of collector and first stage begin to decrease after the tracker system was switched off. Fig. 12 also shows the temperature variations of the water- solvent cycle for the test day.  It is noticed that due to the heat exchange between the fluid in heat exchanger and saline water temperatures of the water-solvent in the trough pipe  inlet, $T_{SC\ inlet}$, and the water-solvent in the heat exchanger,  $T_{H\varepsilon}$ (installed in the bottom of the still) have higher values than the temperature values of the outlet water-solvent from the heat exchanger.

Accumulative fresh water productivities is shown in Figs. 13. It is found that the high quantities of fresh water are obtained from the system while there is no vacuuming of still and just with using a parabolic trough.

Since the trays were made of aluminum with 1.5 mm thickness with high thermal conductivity Experimental data acquired shows no considerable difference in the temperatures on the sides of surface of each tray.

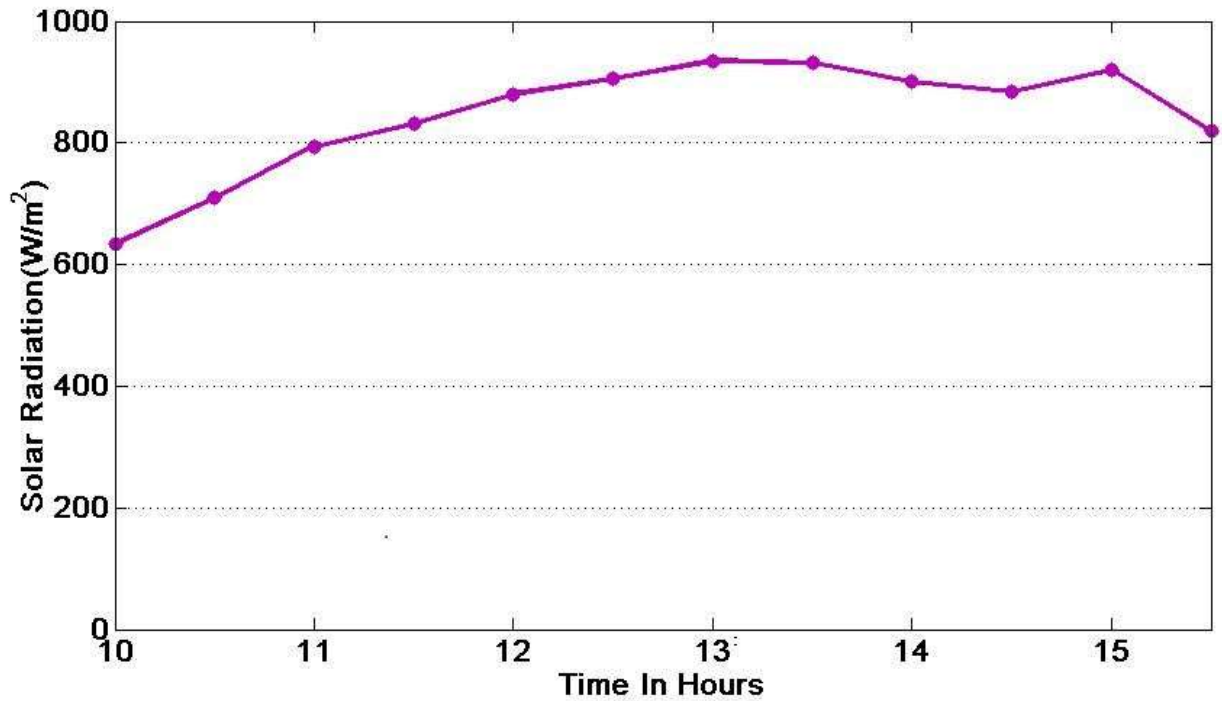

Fig.10. Radiation on the 30 ° surface collected in test region in July 8th 2010

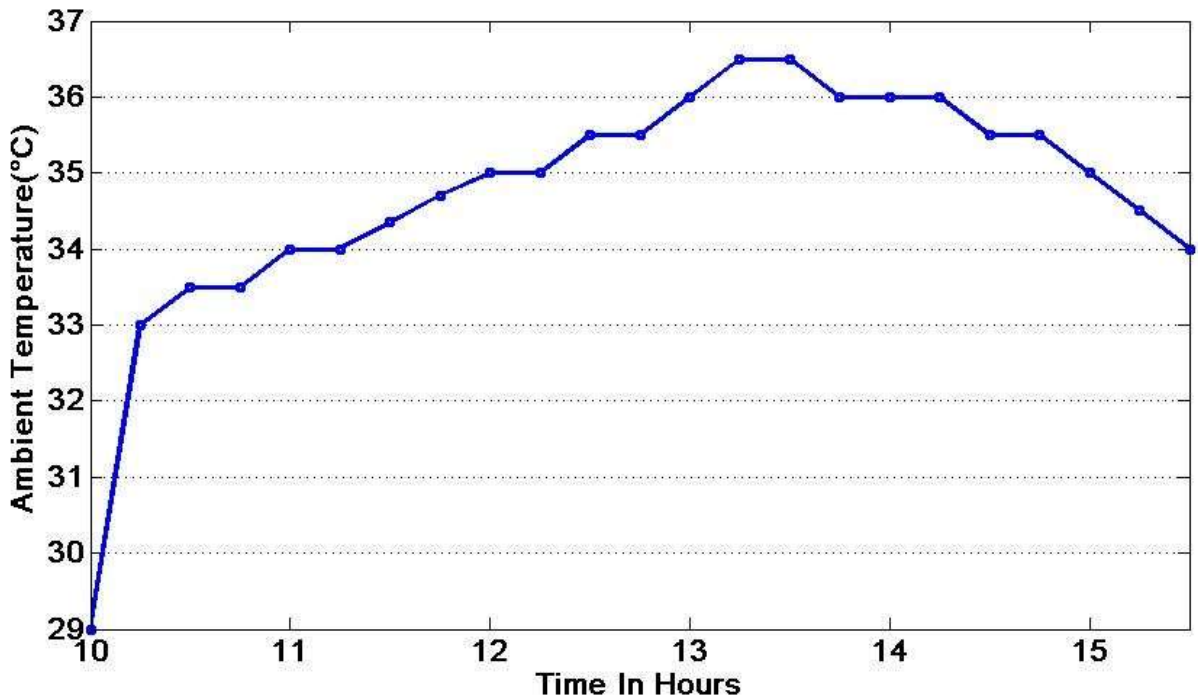

Fig.11.Ambient temperature in the test day

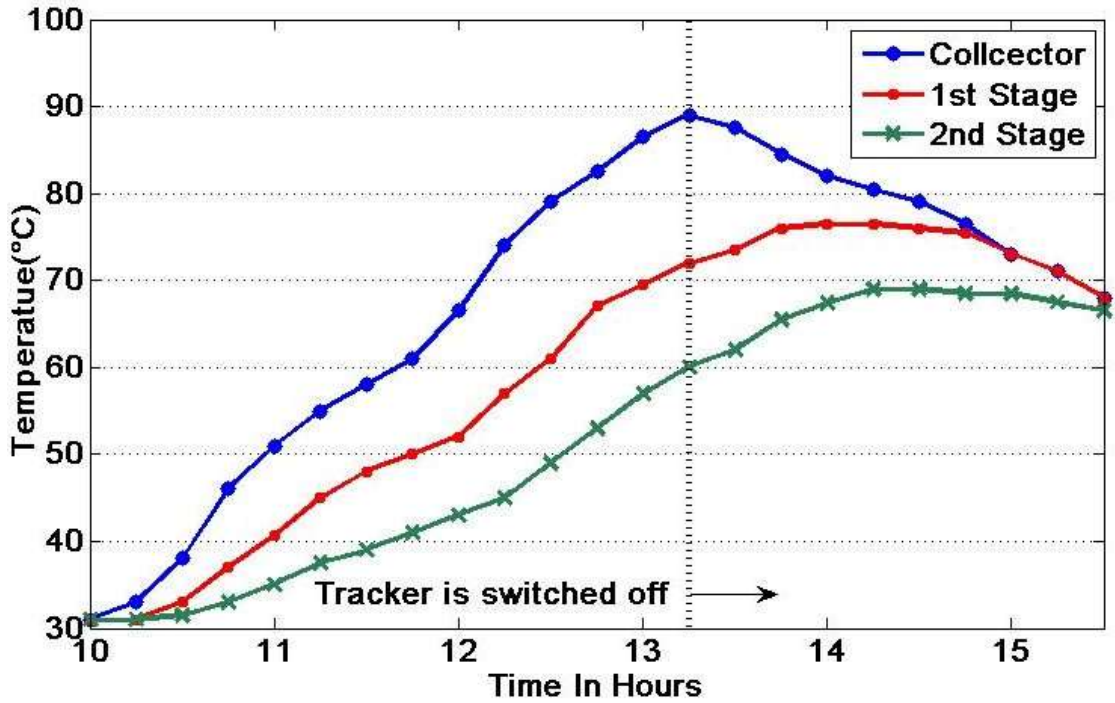

Fig.12. Hourly variation of temperature in solar collector and stages

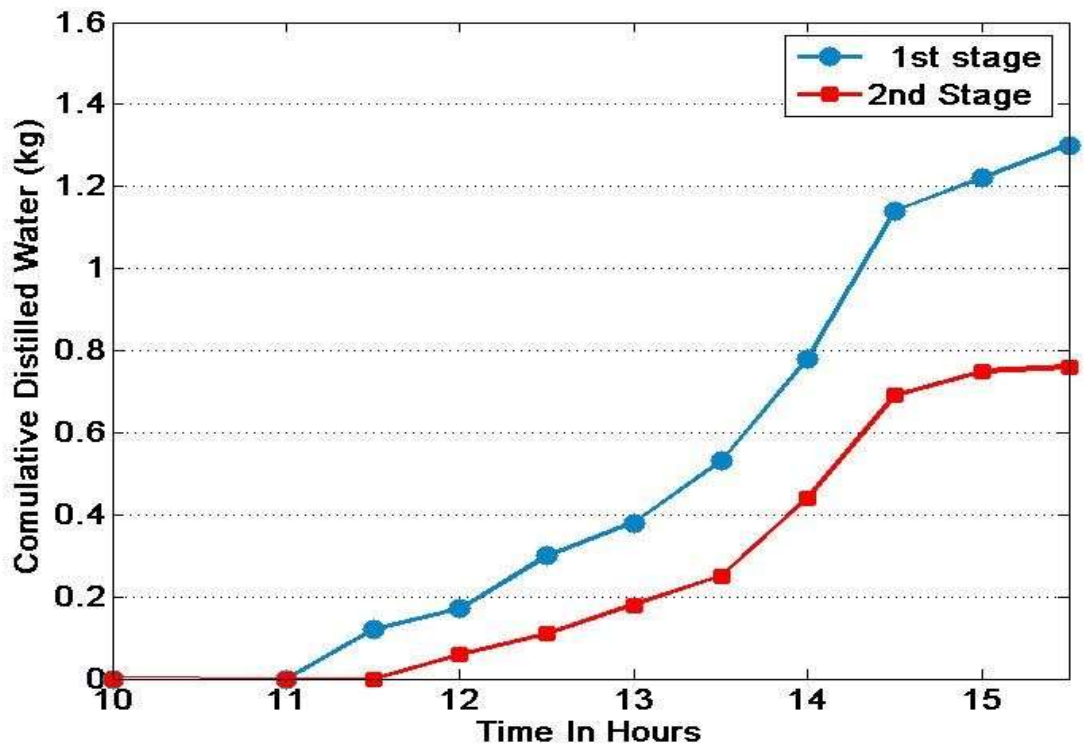

Fig.13.Hourly accumulative fresh water production

## 6. Conclusion

Experimental investigation of a multistage solar still using a parabolic trough is presented. Experimental results showed a very good output compared to the ordinary basin type. The fresh water production capacity of the investigated ''two-stage still with solar parabolic trough'' system was found to be about 3.6 kg/$m^2$ which is for only 5 hours i.e. from 10 A.M. to 3:00 P.M. and it will be more if the solar collector tracks the sun for all day long. The production is higher than that of conventional basin stills and even multi-stage stills coupled with flat plate collectors. The distillation capacity can be significantly increased since the tested system is just a two-stage laboratory prototype. High production can also be reached if the numbers of stages increased since each stage begins to produce distilled water when the temperature reached more than 42 degrees. In presented solar still there is no need to evacuate the still in order to reach high temperatures since evacuating the still is not easy and needs a great attention. Furthermore, in an evacuated still all the distilled water receivers should be evacuated and this makes the system complex, but by means of a parabolic solar trough which is controlled with a sun tracker, high temperatures can be easily reached since these solar collectors have high efficiency and the collector is faced to sun all day long.

## 7. Acknowledgements

The authors would like to thank the International Center for Science & High Technology & Environmental Science of Mahan for funding this project. In addition, thanks are due to Mr. Momenaee for his help in fabrication of apparatus.